\documentclass{aa}
\usepackage{epsfig}

\begin{document}

%   \thesaurus{08.09.2($\theta$Hya); 08.02.3; 08.23.1}
    \thesaurus{08.09.2(16 Dra); 08.02.3; 08.23.1}
   \title{Spectroscopic confirmation of a white dwarf companion to the 
B star 16 Dra}

   \author{M. R. Burleigh \and M. A. Barstow}

   \offprints{Matt Burleigh, mbu@star.le.ac.uk}

   \institute{Department of Physics and Astronomy, University of
                Leicester,
                Leicester LE1 7RH, UK }

   \date{Accepted 22nd May 2000}

    \titlerunning{A white dwarf companion to 16 Dra} 

    \authorrunning{M. R. Burleigh \and M. A. Barstow }

\maketitle

\begin{abstract}

Using an Extreme Ultraviolet Explorer (EUVE) 
spectrum, we confirm the identification 
of a white dwarf companion to the B9.5V star 16 Dra (HD150100), and 
constrain its 
surface temperature to lie between 29,000K and 35,000K. This is the third 
B star $+$ white dwarf non-interacting Sirius-type binary to be 
confirmed, after y Pup (HR2875, HD59635) and $\theta$ Hya (HR3665, HD79469). 
16 Dra and its white dwarf companion are members of a larger resolved 
proper motion system including 
the B9V star 17 Dra A (HD150117). The white dwarf must have 
evolved from a progenitor more massive than this star, 
$M_{\rm MS}\approx3.7M_\odot$. 
White dwarf companions to B stars are important since they set an 
observational limit on the maximum mass for white dwarf progenitors, and can 
potentially be used to investigate the 
high mass ends of the initial-final mass 
relation and the white dwarf mass-radius relation.

   \end{abstract}

%
%  14.Sep.'90: Demo-Vs.
%________________________________________________________________

\section{Introduction}

Unresolved Sirius-type binary systems  
consisting of a white dwarf and a main 
sequence star (spectral type B$-$K) 
are difficult to identify optically, since the bright  
main sequence companion completely swamps the degenerate star's flux. 
However, through the ROSAT Wide Field Camera (WFC, Pounds et~al. 1993) 
and Extreme Ultraviolet Explorer (EUVE, Bowyer et~al. 1994) surveys, 
EUV radiation with the spectral signature of a hot white dwarf has been 
detected originating from apparently inactive 
main sequence stars, giving a clue 
to the existence of a previously unidentified population of Sirius-type 
binaries. Over 20 new systems have now been identified (e.g.~Barstow et~al. 
1994, Burleigh et~al. 1997, Vennes et~al. 1998). For companions of spectral 
type $\sim$A5 or later, far-ultraviolet spectra obtained with the 
International Ultraviolet Explorer (IUE) have been used to confirm the 
identifications, since the white dwarf is actually the brighter 
component at these wavelengths. 
Unfortunately, stars of spectral types O, B and early A will still dominate 
any emission from a white dwarf in the far-UV regime, and IUE or HST 
cannot be used to identify any putative 
degenerate companions to these objects.  

\begin{figure}
\vspace{9cm}
\includegraphics{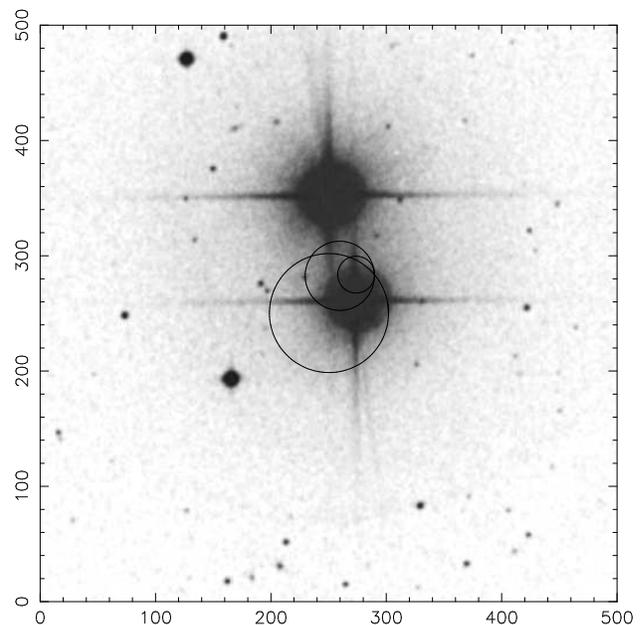}
\caption{Optical image of the 16 Dra system from the Digitized Sky Survey. 
The field is 8$\times$8 arcmin. The lower of the bright stars is 16 Dra. 
The upper bright star is the 17 Dra A/B pair, unresolved here but in fact 
separated by 3.2 arcsec. The circles denote, in decreasing order 
of size, the ROSAT WFC, EUVE and ROSAT PSPC source error boxes.}
\end{figure}

\begin{figure*}
\vspace{8cm}
\includegraphics{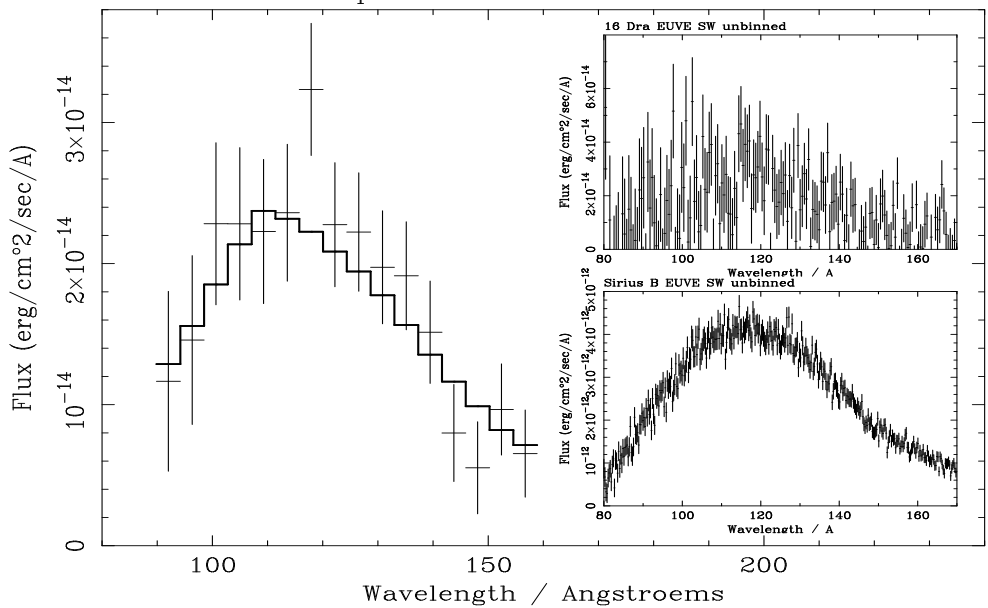}
\caption{EUVE short wavelength spectrum of 16 Dra, shown 
together with a pure-H white dwarf $+$ ISM model (solid line)  
for log $g=8.5$, $T_{\rm eff}=31,900$K, 
$N_{\rm HI}=8.2\times10^{18}$ atoms cm$^{-2}$, 
$N_{\rm HeI}=9\times10^{17}$ 
atoms cm$^{-2}$, and $N_{\rm HeII}=3\times10^{17}$ atoms cm$^{-2}$. Inset 
(upper): the same data, displayed at the instrument resolution of 
$\approx0.5${\AA}. Inset (lower): EUVE short wavelength spectrum of 
Sirius~B, for comparison. }
\end{figure*}

Two bright B stars, 
$\theta$ Hya (HR3665) and y Pup (HR2875), 
were unexpectedly detected in the 
ROSAT and EUVE surveys. Since their soft X-ray and EUV colours were similar 
to many known hot white dwarfs, it was suspected that they too were hiding 
hot white dwarf companions. Fortunately, both EUV sources were bright 
enough to be observed by EUVE's spectrometers. y Pup was observed by EUVE  
in 1996, $\theta$ Hya in 1998, and the formal 
discovery of these Sirius-type systems was subsequently 
reported by Vennes et~al.  
(1997, y Pup), Burleigh \& Barstow (1998, y Pup), and 
Burleigh \& Barstow (1999, $\theta$ Hya). 

White dwarf companions to B stars are of significant importance since they 
must have evolved from massive progenitors, perhaps close to the maximum 
mass for white dwarf progenitor stars. They are also likely to be 
significantly more massive than the mean for white dwarf stars in general 
($\approx0.57M_\odot$, Bergeron et~al. 1992). 
The value of the maximum mass for a 
white dwarf progenitor star, and hence the minimum mass for producing a 
Type II supernova through core collapse in a single star, 
is a long-standing astrophysical 
problem. Weidemann (1987) gives the limit as $\sim8M_\odot$ in 
his semi-empirical initial-final mass relation. Observationally, 
this limit is best set by the white dwarf companion to y Pup (HR2875). 
Echelle spectroscopy of this object by Vennes (2000) 
has recently revealed that this system comprises 
two main sequence B stars (B3.5V$+$B6V) 
in an eccentric $\approx15$ day orbit, with the white dwarf 
forming a third, wider component. The white dwarf must then 
have evolved from a star more massive than B3.5V, $\sim5.5M_\odot$ 
(Vennes 2000). We also note that Bergh\"ofer et~al. 
(2000) have recently suggested that the spectroscopic companion to the B1.5IV 
star $\lambda$ Sco might be a hot ultramassive white dwarf 
($1.25M_\odot<M_{\rm WD}<1.4M_\odot$), based on an excess of EUV and soft 
X-ray radiation detected in ROSAT and EUVE photometric observations. If the 
existence of a white dwarf in this system was confirmed it would obviously 
set the lower limit on the maximum mass for white dwarf progenitors at a 
value near Weidemann's semi-empirical $\sim8M_\odot$ limit, although 
unfortunately $\lambda$ Sco is a close binary ($P=5.959$ days) and mass 
transfer may have taken place at some stage. 

\begin{table*}
\begin{center}
\caption{X-ray and EUV count rates (counts/ksec) and 
size of source error box (radius in arcsec). 
The EUVE source error box is a nominal 30 arcsec, since 
no value is given in Bowyer et~al. (1994). The ROSAT WFC error includes 
the formal 90\% error and a quadratically added 30 arcsec error to account 
for systematics in the aspect solution. }
\begin{tabular}{llccccccccc}
 &  & WFC & & & PSPC & & & EUVE & &  \\
ROSAT No. & Name & error box & S1 & S2 & error box & 0.1-0.4~keV & 
0.4-2.4~keV & error box & 100\AA & 200\AA \\
% & & (arcsec) & (counts/ksec) & (counts/ksec) & (arcsec) &  (counts/ksec) 
%& (counts/ksec) & (arcsec) & (counts/ksec) & (counts/ksec)  \\
RE~J1636$+$525 & 16 Dra & 51.6 & 12$\pm$4 & 46$\pm$11 & 16 & 72$\pm$15 & 0.0 
& 30 & 32$\pm$4 & 0.0 \\ 
\end {tabular}
\end{center}
\end{table*}

% which must 
%have evolved from a progenitor more massive than B5 ($6-6.5M_\odot$).

White dwarfs in Sirius-type binaries can also be used to 
investigate the relationship between the 
mass of a main sequence star and its white dwarf progeny,  
the initial-final mass relation. In particular, white dwarf $+$ B star  
binaries can be used to 
investigate the upper end of this relation. Likewise, 
if the two components can eventually be resolved and an astrometric 
mass determined for the white dwarf, these systems can 
potentially be used to investigate the high mass end of 
the theoretical white dwarf mass-radius relation, for which few observational 
data points currently exist (Vauclair et~al. 1997, Provencal et~al. 1998). 

In addition to y Pup and $\theta$ Hya, another B star was also detected in the 
ROSAT WFC and EUVE surveys, 16 Dra (HD150100, B9.5V). In this paper we 
present an EUVE spectrum of 16 Dra, which proves that it too has a hot white 
dwarf companion. 

%Sirius-type binaries and ROSAT, B star binaries, importance: maximum mass, 
%inital-final mass relation, mass-radius relation for massive WDs (iron cores?)

\section{The 16 Dra system}

16 Dra (HR6184, HD150100, ADS10129C) is a V=5.51 B9.5V star in a visual triple 
system with two other early-type stars, 17 Dra A (HR6185, HD150117, ADS10129A, 
B9V, V=5.08) and 17 Dra B (HR6186, HD150118, ADS10129B, A1V, V=6.58). 
The Hipparcos Catalogue (Perryman et~al. 1997) gives the separation of 17 Dra 
A\&B as just 3.208 arcsec; 17 Dra A and 
16 Dra are then separated by 90.17 arcsec (Fig.1). 
The Hipparcos parallax for 16 Dra is 8.16$\pm$0.55~mas, corresponding 
to a distance of 122.5~pc (114.8$-$131.4~pc), and the parallax for 17 Dra A 
is similar,  
8.22$\pm$0.60~mas, corresponding to a distance of 121.6~pc 
(113.4$-$131.2~pc). 
No solution is given for 17 Dra B. The proper motions of 16 Dra and 
17 Dra A are also very similar: for 16 Dra 
$\mu$$_\alpha$$=$$-$12.9$\pm$0.6~mas/yr  
and $\mu$$_\delta$$=$28.7$\pm$0.6~mas/yr; for 17 Dra A  
$\mu$$_\alpha$$=$$-$12.3$\pm$0.6~mas/yr and 
\newline 
$\mu$$_\delta$$=$27.4$\pm$0.6~mas/yr. All three stars are therefore almost 
certainly related. In that case, any white dwarf in the system must have 
descended from a progenitor more massive than the earliest-type star 
extant, B9V (17 Dra~A).

%16 Dra itself and its neighbours, one of which is B9V, any white dwarf in the 
%system must have come from a more massive progenitor

\section{Detection of EUV radiation from 16 Dra in the ROSAT WFC and EUVE 
surveys}

The ROSAT EUV and X-ray all-sky surveys were conducted between July 1990 and 
January 1991; the mission and instruments are described elsewhere 
(e.g.~Tr\"umper 1992, Sims et~al. 1990). 16 Dra is associated with 
the WFC source RE~J1636$+$525, and was later also detected in the EUVE all-sky 
survey (conducted between July 1992 and January 1993). This source is 
also coincident with a ROSAT Position Sensitive Proportional Counter (PSPC) 
soft X-ray detection. The count rates from all three instruments and 
associated filters are given in Table 1. The WFC count rates are taken from 
the revised 2RE Catalogue (Pye et~al. 1995), which was constructed using 
improved methods for source detection and background screening. The EUVE 
count rates are taken from the First EUVE Source Catalog 
(Bowyer et~al. 1994). The source 
is not included in the revised Second EUVE Source 
Catalog (Bowyer et~al. 1996, see discussion below). 
The PSPC count rate was obtained from the on-line 
ROSAT All Sky Survey Bright Source Catalogue source browser, 
maintained by the Max Planck Institute in Germany 
(Voges et~al. 1999)\footnote{http://www.rosat.mpe-garching.mpg.de/cats/src-browser/}.

Fig.1 shows an optical image of the 16 Dra field from the Digitized 
Sky Survey\footnote{http://ledas-www.star.le.ac.uk/DSSimage/}, 
including the nearby pair 17 Dra A/B. Also shown are the ROSAT WFC, EUVE 
and PSPC source error boxes. Clearly, no obvious optical counterpart 
is visible or resolved within the intersection of 
these three boxes, other than 16 Dra itself. 

The EUV and soft X-ray colours are similar to known hot white dwarfs, and 
the EUV radiation is too strong for it to be the result of UV leakage 
through the WFC filters, although in the EUVE source catalogs (e.g.~Bowyer 
et~al. 1994) it is flagged as such (and, indeed, omitted from the 
second EUVE source catalog as a result, Bowyer et~al. 1996). Far-UV 
leakage is a known problem for EUVE, but the effect is almost 
negligible in the WFC, especially for a late B star like 16 Dra. 
Assuming a temperature of 10,000K, we estimate the far-UV leakage contribution 
to the WFC S2 flux at just 3$\times$10$^{-5}$ counts/sec, compared with the 
$\approx$0.05 counts/sec detected. 
Add the fact that the EUV detection 
is also coincident with a PSPC soft X-ray detection, and it can be safely 
assumed that it is real. 
     
16 Dra is only 
detected in the soft 0.1$-$0.4 keV PSPC band; only one (rather unusual) 
white dwarf has ever been detected at higher energies (KPD0005$+$5105, 
Fleming et~al. 1993). Most active stars are also hard X-ray sources, and 
indeed Bergh\"ofer et~al. (1996) found only three of the B stars detected 
by the ROSAT PSPC were not hard X-ray sources. Interestingly, these 
are y Pup, $\theta$ Hya (both of which have confirmed white dwarf 
companions) and 16 Dra. Therefore, Burleigh \& Barstow (1999) confidently 
predicted that 16 Dra might also be hiding a hot white dwarf companion, 
and using the ROSAT count rates demonstrated that it probably had a surface 
temperature between 25,000$-$37,000K.

%same as for theta Hya, plus Fig.of stars and surveys error boxes

\section{EUVE pointed observation and data reduction}

16 Dra was observed twice by EUVE in dither mode, firstly for 
$\approx$220,000 sec between 1999 February 28th and 1999 March 7th, 
and then for $\approx$230,000 sec between 1999 June 27th and 1999 
July 6th, giving a total exposure time of 453,985.5 sec. We have extracted 
the spectra from the detector images using standard IRAF procedures. 
Our general reduction techniques are described in earlier work (e.g.~Barstow 
et~al. 1997).

The target was only detected, weakly, in the short (70$-$190{\AA}) 
wavelength spectrometer. To improve the signal/noise ratio, 
we combined the two 
observations before extracting this spectrum, and then binned the data 
by a factor 32. The resultant spectrum is shown in Fig.2. The flux 
distribution is similar in shape to the familiar EUV continuum expected 
from hot white dwarfs in this spectral region (Fig.2, inset). 
No emission lines are visible in the raw data (Fig.2, inset), 
despite the apparent excess of flux at $\sim$120{\AA} in the binned 
spectrum.

The only stars other than white dwarfs whose photospheric EUV 
radiation has been detected by the ROSAT WFC and EUVE are the bright 
early B giants $\beta$ CMa (B1II$-$III, Cassinelli et~al. 1996) and 
$\epsilon$ CMa (B2II, Cohen et~al. 1996). The photospheric continuum of 
$\epsilon$ CMa is visible down to $\sim$300{\AA}, although no continuum 
flux from $\beta$ CMa is visible below the HeI edge at 504{\AA}. 
Both stars also have strong EUV and X-ray emitting winds, and in $\epsilon$ 
CMa emission lines are seen by EUVE in the short and medium wavelength 
spectrometers from e.g. high ionisation features of iron. 
Similarly, strong narrow emission features of e.g. oxygen, nickel and 
calcium are commonly seen in EUV spectra of active stars and RS CVn systems. 
Since no such features are visible in this spectrum of 16 Dra, 
we can categorically rule out a hot wind or unresolved active late-type
companion to 16 Dra as an alternative source for the EUV radiation. 

%as theta Hya, add binning factor, discuss B star detections by ROSAT and only 
%these three have soft spectra, discuss UV transmission (lack of with ROSAT), 
%compare Beta CMa, no emission lines, excess in one data point not 
%indicative of an emission feature 
%Fig.-\begin{table*}
%Further figure - EUV spectrum, IUE spectrum, optical spectrum, plus WD model 

\section{Analysis of the hot white dwarf's EUV spectrum}

We have matched the EUV spectrum of 16 Dra with a grid of hot white dwarf 
$+$ ISM model atmospheres, in order to constrain the atmospheric 
parameters (temperature and surface gravity) of the degenerate star and 
the interstellar column densities of {\small 
HI, HeI} and  {\small HeII}. Unfortunately there 
are no spectral features in this wavelength region to give us an 
unambiguous determination of $T_{\rm eff}$ and log~$g$. However, by making 
a range of assumptions to reduce the number of free parameters in 
our models, we can place constraints on some of the white dwarf's physical 
parameters. Our method is similar to that used in the analysis of y Pup 
(Burleigh \& Barstow 1998) and $\theta$ Hya (Burleigh \& Barstow 1999).

\begin{figure}
\vspace{7cm}
\includegraphics{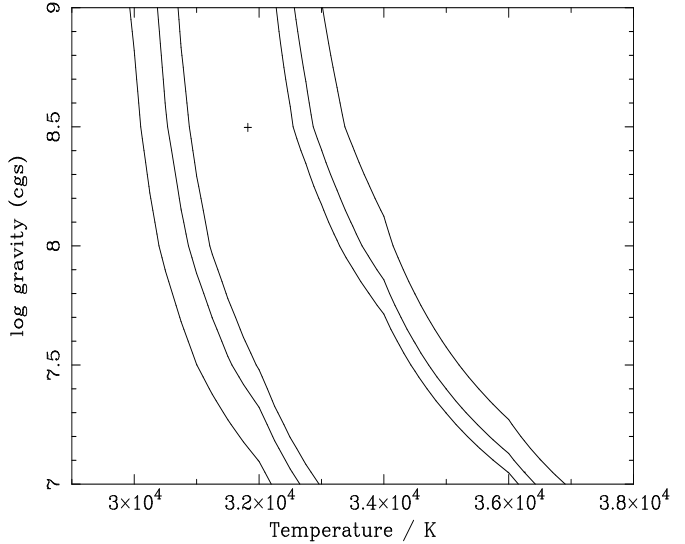}
\caption{Confidence contours for the spectral fit to the EUVE data, at 66\%, 
90\% and 99\% in the ($T_{\rm eff}$, log~$g$) plane.}
\end{figure}

\begin{figure}
\vspace{7cm}
\includegraphics{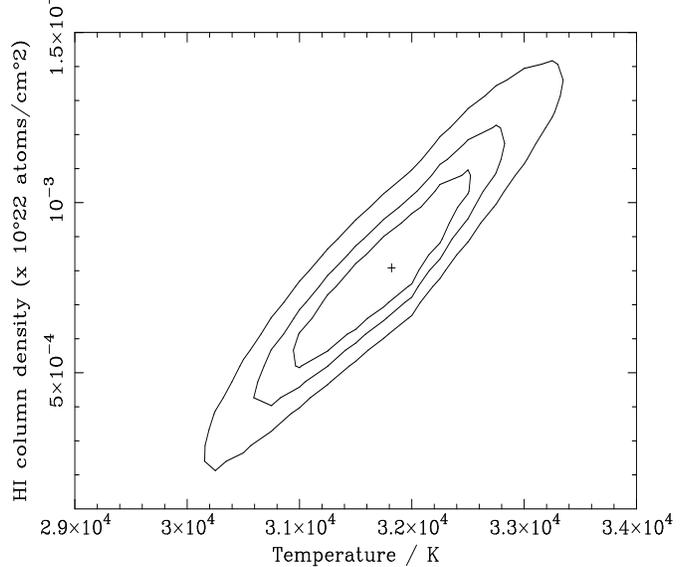}
\caption{Confidence contours for the spectral fit to the EUVE data, at 66\%, 
90\% and 99\% in the (log~$g$, $N_{\rm HI}$) plane.
}
\end{figure}

\begin{table}
\begin{center}
\caption{Hamada-Salpeter zero-temperature mass-radius relation. 
16 Dra is assumed to lie at a distance of 122.5pc.}
\begin{tabular}{ccccc}
log $g$ & $M_{\rm WD}$ & $R_{\rm WD}$ & $R_{\rm WD}$ & $(R_{\rm WD}/D)^2$ \\
 & ($M_\odot$) & ($R_\odot$) & ($\times10^8$cm) & \\
7.5 & 0.30 & 0.017 & 11.832 & 9.929$\times$10$^{-24}$ \\
8.0 & 0.55 & 0.013 & 9.048 & 5.806$\times$10$^{-24}$ \\
8.5 & 0.83 & 0.009 & 6.264 & 2.783$\times$10$^{-24}$ \\
9.0 & 1.18 & 0.006 & 4.176 & 1.237$\times$10$^{-24}$ \\
\end {tabular}
\end{center}
\end{table}

\begin{table*}
\begin{center}
\caption{WD parameters and interstellar column
densities.}
\begin{tabular}{ccccccc}
log~$g$ & $T_{\rm eff}$ (K) & 90\% range & $N_{\rm HI}$  
& 90\% range & $N_{\rm HeI}$ & $N_{\rm HeII}$ \\
 & & & $\times$10$^{19}$ & $\times$10$^{19}$ & $\times$10$^{18}$ & 
$\times$10$^{17}$ \\
7.5 & 29,300 & (28,100$-$30,400) & 1.0 & (0.4$-$1.5) & 1.1 & 4.1 \\
8.0 & 30,100 & (28,900$-$31,200) & 0.9 & (0.4$-$1.4) & 1.0 & 3.7 \\
8.5 & 31,900 & (30,500$-$33,000) & 0.8 & (0.4$-$1.2) & 0.9 & 3.4 \\
9.0 & 34,200 & (32,800$-$35,400) & 0.7 & (0.3$-$1.1) & 0.8 & 3.0 \\
\end {tabular}
\end{center}
\end{table*}

Firstly, we assume that the white dwarf has a pure-hydrogen atmosphere. 
This is a reasonable assumption to make, since Barstow et~al. (1993) 
first showed that for T$_{\rm eff}<40,000$K hot DA 
white dwarfs have an essentially 
pure-H atmospheric composition. We can then fit a range of models, each fixed 
at a discrete 
value of the surface gravity log~$g$. However, before we can do this 
we need to know the normalisation parameter of each model, which is 
equivalent to $(R_{\rm WD}/D)^2$. For this, 
we can use the Hipparcos 
parallax to give us the distance, 
and the Hamada \& Salpeter (1961) zero-temperature 
mass-radius relation to give us the radius of the white dwarf 
corresponding to each value of the surface gravity (see Table 2).

We can also reduce the number of unknown free parameters in the ISM model. 
From EUVE spectroscopy, Barstow et~al. (1997) measured the line-of-sight 
interstellar column densities of {\small HI, HeI} and 
{\small HeII} to a number of hot 
white dwarfs. They found that the mean {\small H} 
ionisation fraction in the local 
ISM was 0.35$\pm$0.1, and the mean {\small He} fraction was 0.27$\pm$0.04. 
From these estimates, and assuming a cosmic {\small H/He} 
abundance, we calculate 
the ratio $N_{\rm HI}/N_{\rm HeI}$ in the local ISM$=$8.9 and 
$N_{\rm HeI}/N_{\rm HeII}=2.7$. 
We can then fix these column density ratios in 
our model, leaving us with just two free parameters - temperature and the 
{\small HI} column density.

The model fits at a range of surface gravities from log~$g=7.5-9.0$ are 
summarized in Table 3. Note that our range of fitted temperatures is in 
agreement with those of Burleigh \& Barstow (1999), who modelled the ROSAT 
EUV and soft X-ray photometric data for 16 Dra on the assumption that the 
source was indeed an unresolved white dwarf.

%as theta Hya
%Tables: H-S M-R relation, fitted parameters & column densities

\section{Discussion}

\begin{table*}
\begin{center}
\caption{The earliest-type stars known to have white dwarf companions}
\begin{tabular}{llcccccl}
Name & Alt. name & Sp. Type & $M_{\rm MS}$ & Period & Distance (pc) & 
$M_{\rm WD}$ & Reference \\
 & & & ($M_\odot$) & & (Hipparcos) &  ($M_\odot$) & \\ 
$\lambda$ Sco$^\star$ & HR6527 & B1.5IV & $\sim$9 & 5.959 days &  
180$-$265 & 1.25$-$1.4 & Bergh\"ofer et~al. (2000) \\
y Pup & HR2875 & B3.5V$+$B6V & 5.5 & & 157$-$187 & $>$0.91 & 
Burleigh \& Barstow (1998) \\
 & & & & & & & Vennes et~al. (2000) \\
16 Dra$^\dagger$ & HD150100 & B9V & 3.7 & & 115$-$131 & $>$0.69 & 
this paper \\
$\theta$ Hya$^\dagger$$^\dagger$ & HR3665 & B9.5V & 3.4 & $\buildrel > \over 
{_\sim}$10 yrs & 38$-$41 & $>$0.68 & Burleigh \& Barstow (1999)  \\
Sirius B$^\ddagger$ & & A0V & 3.25 & $\approx$50 yrs & 
2.637$\pm$0.011 & 1.034$\pm$0.026 & Holberg et~al. (1998) \\
Beta Crt.$^\dagger$$^\dagger$ & & A1III & 2.9 & 
$\buildrel > \over {_\sim}$10 yrs$$ & 77$-$87 & 0.44 & 
Vennes et~al. (1998) \\  
\end {tabular}
\end{center}
Main sequence masses from Allen (1973). \\
$\star$ Suggested, not confirmed. Note that the Hipparcos distance estimate  
is inconsistent with the photometric distance ($\sim140$pc). \\ 
An alternative explanation is that the system consists of two B stars. \\
$^\dagger$ The spectral type of 16 Dra is in fact 
B9.5V. Its proper motion companion 
17 Dra A is B9V, and thus any white dwarf in the system must have evolved from 
a progenitor more massive than this. See text for futher discussion.\\
$^\dagger$$^\dagger$ 
Micro-variability in the proper motions of these stars as measured by 
Hipparcos indicate binary periods $\buildrel > \over {_\sim}$10 yrs. \\
$^\ddagger$ Holberg et~al. (1998) use an initial-final mass  
relation to estimate the progenitor mass of Sirius B as 6$-$7M$_\odot$. \\
\end{table*}

We have analysed the weak EUVE spectrum of the B9.5V star 16 Dra, and 
confirm that there is an unresolved hot white dwarf in the field. 

Fig.1 clearly shows that the white dwarf is not resolved from 16 Dra
in the Digitized Sky Survey image, and 
their angular separation can be no more than $\approx$30 arcsec. 
However, if the white dwarf lies at the same distance as 16 Dra and its 
proper motion companions 17 Dra A \& B, and is related to them, then it 
must have evolved from a progenitor more massive than the earliest extant 
star in the system, 17 Dra A (B9V). Thus this degenerate has the second 
most massive progenitor among known white dwarfs. Table 4 lists the 
earliest type stars known to have white dwarf companions, including all 
three B star $+$ white dwarf binaries and Sirius. 

EUVE spectra provide us with little information with which to 
constrain a white dwarf's surface gravity, and hence its mass, but we can 
use a theoretical initial-final mass relation between main sequence stars 
and white dwarfs, e.g.~that of Wood (1992), to estimate the mass  of the   
white dwarf if the progenitor was slightly more massive than a B9V star:
$M_{\rm WD} \, = \, A \, exp \, (B \times M_{\rm MS})$,  
where $A \, = \, 0.49M_\odot$ and $B \, = \, 0.094M_\odot^{-1}$. 

For $M_{\rm MS} \, = \, 3.7M_\odot$, we find $M_{\rm WD} \, = \, 0.69M_\odot$.
This would suggest the surface gravity of the white dwarf log $g>8.0$ and, 
therefore, its surface temperature most likely lies between $\approx29,000$K 
and $\approx35,000$K. 

Finally, we note that if this white dwarf can be resolved from 16 Dra, 
then an optical spectrum may potentially be obtained (e.g.~with HST/STIS) 
from which its temperature and gravity can be tightly constrained. The mass 
can then be estimated, and this binary could be used to investigate 
the initial-final mass relation and to test the high mass end of the 
mass-radius relation.

\begin{figure*}
\vspace{8cm}
\includegraphics{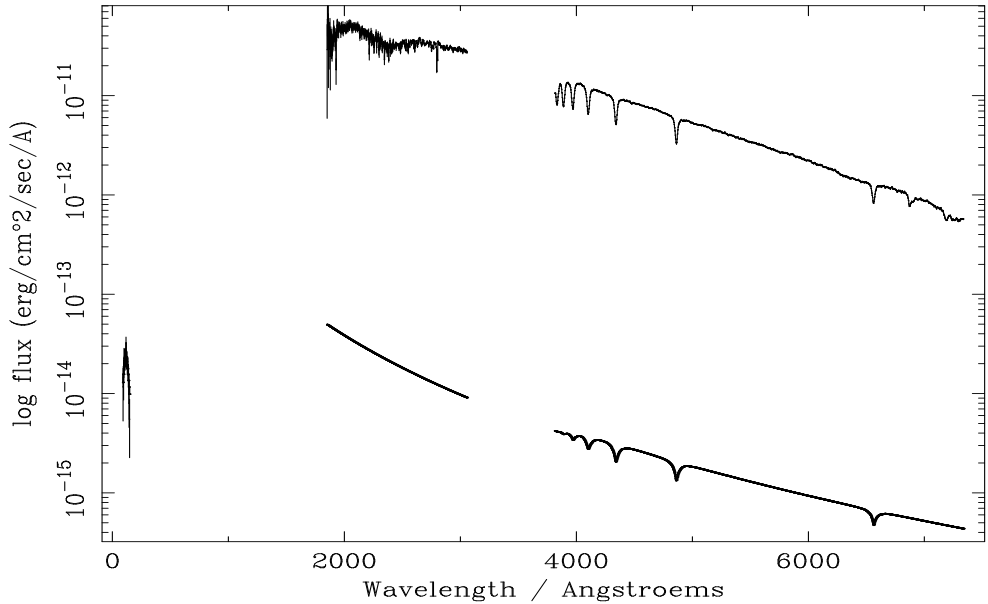}
\caption{Left to right: EUV, UV and optical spectra of 16 Dra, shown 
together with a white dwarf model for $T_{\rm eff}=31,900$K and log $g=8.5$. 
The B9.5V star clearly dominates the UV and optical flux, and the 
white dwarf is only detectable in the EUV.  
The UV spectrum was extracted from the IUE final archive (LWP18244). The 
optical spectrum was obtained on 1997 March 17 at the Russain Academy of 
Sciences' Special Astrophysical Observatory's 6m telescope, located in the 
Karachaevo-Cherkesia region of southern Russia. 
   }
\end{figure*}

\begin{acknowledgements}

Matt Burleigh is the UK ROSAT Support Scientist and 
acknowledges the support of PPARC, UK. 
We thank Detlev Koester (Kiel) for the use of his model atmosphere grids, 
and Jurek Madej (Warsaw University) and Victor Bychkov (Special Astrophysical 
Observatory, Russian Academy of Sciences) 
for their help in obtaining and reducing the optical spectrum 
of 16 Dra. This research has made us of the {\it SIMBAD} database operated 
by CDS, Strasbourg, France, and the Leicester Database and Archive Service 
({\it LEDAS}). 

\end{acknowledgements}

\section{Appendix: HD93847 - another B star $+$ white dwarf binary in the 
ROSAT WFC catalogue?}

A fourth B star, HD93847 (B9, V$=$7.46) was detected in the WFC survey, with a 
count rate of 9$\pm$4 counts/ksec in the S1 filter and 18$\pm$5 counts/ksec 
in the S2 filter. Since the S2/S1 count rate ratio is similar to known hot 
white dwarfs, it would be reasonable to suggest that perhaps this B star 
is also hiding a degenerate companion. The detected flux is highly unlikely 
to be due to far-UV leakage: we estimate this contribution as only 
6$\times$10$^{-6}$ counts/sec. Unfortunately, this weak WFC source 
was not detected by the ROSAT PSPC or by EUVE, and thus it is not clear 
whether the detection is real. With a combined S1$+$S2 detection significance 
of only 5.8$\sigma$ it is in the regime where a few spurious detections 
are expected (see Pye et~al. 1995). Alternatively, the B star may be 
hiding an unresolved active late-type companion that flared in the EUV 
waveband. We are therefore unwilling to claim 
that this source is due to another hidden white dwarf.

\end{document}